\documentclass[useAMS,usenatbib]{mn2e}
\usepackage{graphicx}
\title[On minor black holes in galactic nuclei]{On minor black holes in galactic nuclei}

\author[B. McKernan, K.E.S. Ford, T.Yaqoob \& L.M.Winter]{B. McKernan$^{1,2,3}$\thanks{E-mail:bmckernan at amnh.org (BMcK)}, K.E.S. Ford$^{1,2,3}$, T. Yaqoob$^{4}$ \& L.M.Winter$^{5,6}$\\
$^{1}$Department of Science, Borough of Manhattan Community College, City University of New York, New York, NY 10007\\
$^{2}$Department of Astrophysics, American Museum of Natural History, New York, NY 10024\\
$^{3}$Graduate Center, City University of New York, 365 5th Avenue, New York, NY 10016\\
$^{4}$Department of Physics \& Astronomy, Johns Hopkins University, Baltimore, MD 21218\\
$^{5}$Center for Astrophysics \& Space Astronomy, University of Colorado, Boulder, CO 80303\\
$^{6}$Hubble Fellow\\
}
\begin{document}

\date{Accepted. Received; in original form}

\pagerange{\pageref{firstpage}--\pageref{lastpage}} \pubyear{2008}

\maketitle

\label{firstpage}

\begin{abstract}
Small and intermediate mass black holes should be expected in galactic nuclei 
as a result of stellar evolution, minor mergers and gravitational dynamical 
friction. If these minor black holes accrete as X-ray 
binaries or ultra-luminous X-ray sources, and are associated with star 
formation, 
they could account for observations of many low luminosity AGN or LINERs. 
Accreting and inspiralling intermediate mass black holes could provide 
a crucial electromagnetic counterpart to strong gravitational wave signatures,
 allowing tests of strong gravity. Here we 
discuss observational signatures of minor black holes in galactic nuclei and 
we demonstrate that optical line ratios observed in LINERs or transition-type 
objects can be produced by an ionizing radiation field from ULXs. We conclude 
by discussing constraints from existing observations as well as candidates for 
future study.  
\end{abstract}


\section{Introduction}
\label{sec:intro}
The largest black holes in the Universe live in the centers of galaxies 
\citep[e.g.][]{b73}. However, small to intermediate mass (or minor) black 
holes must also 
live in galactic nuclei as a result of stellar evolution, minor mergers and 
gravitational dynamical friction. This has two important consequences. First, 
some low luminosity galactic nuclei may not actually be powered by 
inefficient accretion onto a supermassive black hole. Instead, some galactic 
nuclei may be powered by a number of X-ray binaries (XRBs) or ultra-luminous 
X-ray sources (ULXs) possibly with some star formation around a quiet 
supermassive black hole \citep[e.g.][]{b9,b98}. So, while accretion onto 
supermassive black holes can extend down to tiny fractions of Eddington 
luminosity \citep[e.g.][]{b9,b59}, inferring very weak accretion rates becomes 
dangerous at low X-ray luminosities, since it is possible to get lost 
in the XRB or ULX nuclear noise. Second, inspiralling black 
holes offer an excellent opportunity to study strong gravity via gravitational
 radiation \citep[e.g.][]{b34,b8}. An X-ray counterpart to a source of 
gravitational 
radiation would be extremely important since it would provide crucial 
simultaneous electromagnetic 
information about the predictions of strong gravity. Therefore we should 
investigate the occurrence of accreting, non-supermassive black holes in 
galactic nuclei.

In \citet{b98} we proposed that LINER 2s (LINERs devoid of broad 
$\rm{H}\alpha$ wings) without flat-spectrum compact radio cores may be 
powered by nuclear ULXs and low levels of star formation, rather than 
low luminosity accretion onto supermassive black holes. Here we discuss the 
general phenomenon of minor black holes in galactic nuclei and observational 
consequences if they are accreting. In section \S\ref{sec:sims} we discuss constraints on nuclear ULXs from 
BPT optical line ratio diagrams and simulated ULX 
continua. In section \S\ref{sec:obs} we discuss constraints on the occurrence of 
nuclear ULXs (and XRBs) from existing observations. In section \S\ref{sec:future}, we conclude by outlining future 
observational strategies in the X-ray, optical and IR bands to disentangle signatures of 
small mass black holes in galactic nuclei. We also point out that nuclear ULXs may be present in 
'transition' type objects and that the presence of flat-spectrum 
compact radio cores need not rule out the presence of nuclear ULXs. 

\section{Small \& intermediate mass black holes in galactic nuclei}
\label{sec:ulx}
The mass spectrum of astrophysical black holes in the Universe 
is believed to span around nine orders of magnitude. The largest, 
supermassive, black holes ($M_{BH}\sim 10^{6}-10^{9}M_{\odot}$) live in 
galactic centers \citep[e.g.][]{b73}. X-ray observations of minor 
black holes ($\leq 10^{5}M_{\odot}$) inevitably tend to be biased towards 
accreting black holes lying \emph{far} from the centers of galaxies 
\citep[e.g.][]{b6,b34,b12}. However, minor black holes must naturally occur in
 galactic nuclei as a result
 of stellar evolution ($M_{BH}\leq 20M_{\odot}$) or minor mergers by clusters 
and dwarf galaxies ($10^{2} \leq M_{BH} \leq 10^{5}M_{\odot}$). In our own 
Galaxy, a large number of stellar mass black holes ($\sim 20,000$) are 
expected within $\sim 1$pc of SgrA* 
\citep[e.g.][]{b80,b81} and should dominate the mass density at $< 0.2$pc 
\citep{b82}. \citet{b83} have found four X-ray transients within $\sim 1$pc of
 Sgr A*, with (unobscured) peak X-ray luminosity $>10^{36}$ erg $\rm{s}^{-1}$ 
\citep{b84}. Allied with evidence for young massive stars (3-7Myrs) within 
$\sim 1$pc of SgrA*, this indicates that accreting minor black holes could be 
an important part of emission from galactic nuclei.

If minor black holes in galactic nuclei are accreting, their X-ray luminosities could 
mimic low luminosity AGN and LINERs and mis-classification of these galactic 
nuclei can occur. For example, in M82 an accreting black hole of mass 
($500M_{\odot}<M<10^{6}M_{\odot}$) yielding an X-ray luminosity of 
$L_{x} \sim 10^{40}$ erg $\rm{s}^{-1}$, lies a mere $\sim 180$pc from the 
kinematic center of the galaxy \citep{b49}. The optical line ratios from the 
inner $\sim 200$pc of M82 are consistent with a borderline 
$\rm{H}_{II}$ nucleus/transition object \citep{b16}. If an identical galactic 
nucleus at $>30$Mpc were observed with Chandra (angular resolution $\geq 
0.5^{``}$), the X-ray emission would appear to originate in the galactic 
center and might incorrectly be thought due to inefficient accretion onto the 
central supermassive black hole. Likewise, in M31 there are 33 X-ray point 
sources in the innermost $\sim 450$pc of that galaxy, with individual 
luminosities spanning $L_{x}\sim 
[10^{36},10^{38}]$ erg $\rm{s}^{-1}$ \citep{b28}. If an identical galactic 
nucleus at $\sim 100$Mpc were observed with Chandra, it would have a nuclear 
X-ray luminosity of $L_{x} \sim 10^{39}$ erg $\rm{s}^{-1}$, comparable to that 
of some LINERs.
 
Studies of nearby galaxies, such as M81 ($\sim 3.9$Mpc, 124 X-ray sources, 
\citep{b26}), M83 ($\sim 4$Mpc, 81 X-ray sources \citep{b30}) and M101 
($\sim 7$Mpc, 110 X-ray sources \citep{b31}) encourage us to expect (at the 
very least) multiple moderate luminosity X-ray sources in most galactic 
nuclei. In actively star forming galaxies, 
such as The Antennae (NGC 4038/39) at $\sim 19$Mpc, there are nine ULXs with 
$>10^{39}$ erg $\rm{s}^{-1}$\citep{b24}. X-ray luminosity functions (XLFs) can
 help to disentangle sources of X-ray 
emission in galactic nuclei \citep[e.g.][]{b23}. In M31 large difference in XLFs 
between the inner bulge, outer bulge and disk X-ray point source populations 
may be due to limited statistics, differences in stellar ages, or 
contributions from globular clusters \citep[e.g.][]{b28,b32,b23}. Furthermore,
 since X-ray transients can actually dominate XLFs \citep{b38}, the treatment 
of transient outbursts can make a big difference in expected XLFs \citep{b40}. 

\begin{figure}
\includegraphics[width=3.35in]{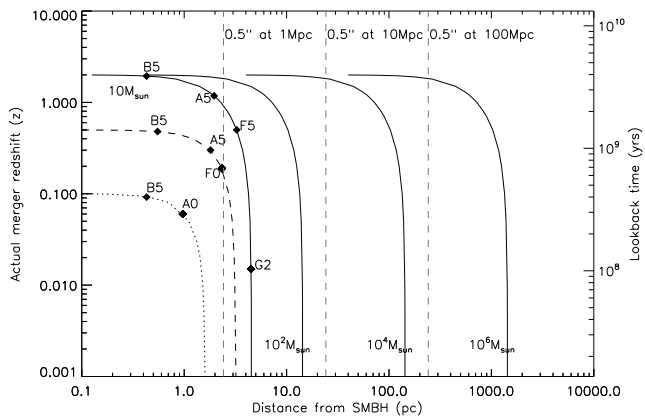}
\caption{Distance of minor black hole from galactic center as a function of 
redshift (left y-axis) and time (right y-axis). Solid curves correspond to
minor black holes of mass $10,10^{2},10^{4},10^{6} M_{\odot}$ formed or 
introduced by minor merger at a redshift of $z=2$. Depending on the 
initial distance (x-axis) at $z=2$, the redshift at which the merger occurs 
lies between $z\sim 2$ and $z\sim 10^{-3}$. Dashed and dotted curves 
correspond to inspiralling $10M_{\odot}$ black holes introduced at $z=0.5$ and
 $z=0.1$ respectively. Vertical dashed lines correspond to Chandra's limiting 
angular resolution of $0.5^{``}$ at 1,10 and 100Mpc (left to right) 
respectively. Filled-in diamonds on $10M_{\odot}$ curves correspond to example
 binary companion main sequence lifetimes. Curves may be read as follows: A 
10$M_{\odot}$ black hole that forms e.g. $\sim2(4)$pc from the central 
supermassive black hole at $z=2$ will merge at $z\sim 1(0.2)$. An A5(F5) 
companion would not last long enough to fuel accretion during the entire 
inspiral to $z=1(0.2)$, but a companion of mass G2 or lower could last this 
long.
\label{fig:z}}
\end{figure}

The orbits of minor black holes in galactic nuclei will tend to decay via 
gravitational dynamical friction \citep[e.g.][]{b1,b34}. So minor black holes 
will tend to migrate deeper into galactic nuclei over time. The dynamical 
frictional time ($t_{fric}$) for an object of mass M at galactic 
radius r, to sink to the galactic center (with central velocity dispersion 
$\sigma$) is given by
\begin{equation}
t_{fric}=\frac{5 \times 10^{9}\rm{yrs}}{\rm{ln} \Lambda}\left(\frac{\rm{r}}{\rm{kpc}}\right)^{2}\frac{\sigma}{200\rm{km/s}}\left(\frac{M}{10^{7}M_{\odot}}\right)^{-1}
\label{eq:dynfric}
\end{equation}
where ln$\Lambda \sim 5-20$ is the Coulomb logarithm \citep[e.g.][]{b34} and a smooth, 
homogeneous background is assumed. Note that this assumption may break down inside $\sim 10$pc,
but we ignore this for simplicity as it does not affect our general argument. 
Figure~\ref{fig:z} shows the distance of minor black holes from a central 
supermassive black hole as a function of redshift (or time). Frictional 
timescales are calculated from equation~\ref{eq:dynfric} assuming ln$\Lambda 
\sim 10$ and $\sigma \sim 200$ km $\rm{s}^{-1}$. 
Solid curves indicate black holes of mass $10,10^{2},10^{4},10^{6}M_{\odot}$ 
formed or introduced by merger at $z\sim 2$. Dashed and dotted curves 
correspond to the inspiral of a $10M_{\odot}$ black hole formed at $z=0.5$ and
 $z=0.1$ respectively. Filled-in circles on the $10M_{\odot}$ curves 
correspond to examples of binary companion main sequence lifetimes. Vertical 
dashed lines in Fig.~\ref{fig:z} correspond to 
the limiting angular resolution of $\sim 0.5^{``}$ on Chandra at 1,10 and 
100Mpc respectively. Thus, for example a $10M_{\odot}$ black hole born 
$\sim 1$pc from the central supermassive black hole at $z\sim 0.1$ will merge 
with the central black hole around $z\sim 0.06$ and could be fuelled during 
the entire inspiral by a binary companion with a mass corresponding to 
spectral class A0 or lower. While $\sim 1$pc seems particularly close to the 
central black hole for star formation/evolution, in our own Galaxy tens of O 
and B stars can be found $0.01-5$pc from the supermassive 
black hole \citep{b56}. In Fig.~\ref{fig:z} the main sequence lifetime of 
binary companions is likely to be the limiting factor for fuelling stellar 
mass black holes ($\leq 20 M_{\odot}$), assuming $\sim 1\%$ Eddington 
accretion and either continuous accretion from O/B companions or sporadic 
outbursts from lower mass companions, as observed in our own Galaxy 
\citep{b71}. For black holes much larger than stellar mass, gravitational 
capture of companions is 
the more likely fuelling scenario, particularly with increasing mass, but this 
is difficult to model appropriately. Nevertheless, some fraction of IMBHs must
 capture companions during their inspiral.

\section{Where should we look?}
\label{sec:preds}
While several accreting black holes should be expected in most galactic 
nuclei, it should be easier to isolate them in certain kinds of nuclei. 
In \citet{b98}, we suggested that certain types of LINER nuclei may in 
fact be powered by nuclear ULXs or XRBs. Evidently, if objects such as M82 and 
M31 were observed from large distances, they would look similar in the X-ray 
band to low luminosity AGN and LINERs observed at these distances 
\citep[e.g.][]{b22}. But what about the optical emission lines from these 
nuclei? In the optical band, diagnostic diagrams, using common, 
low-ionization lines are used to separate galactic nuclei according to 
classification \citep[e.g.][]{b42,b43,b44}. For 
Seyfert AGN, the ionization parameter spans a wide range and can be very high 
\citep[e.g.][]{b97}, placing AGN in the upper right section of diagnostic 
diagrams. LINERs lie below and to the right of a theoretical mixing line, in 
a region where shocks or a power-law ionizing radiation field with low 
ionization parameter can yield the appropriate optical line ratios 
\citep[see e.g.][]{b44}. A nucleus powered by ULXs or multiple XRBs could have
 a low ionization parameter, but a relatively hard (X-ray dominated) radiation 
field which could drive optical line ratios into the 'LINER' section of a 
diagnostic diagram, even though the
 central supermassive black hole is quiescent. Low luminosity galactic 
nuclei may therefore be fruitful places to search for minor nuclear black 
holes.

\subsection{Simulated ULXs powering LINERs}
\label{sec:sims}
\begin{figure}
\includegraphics[width=3.35in,angle=0]{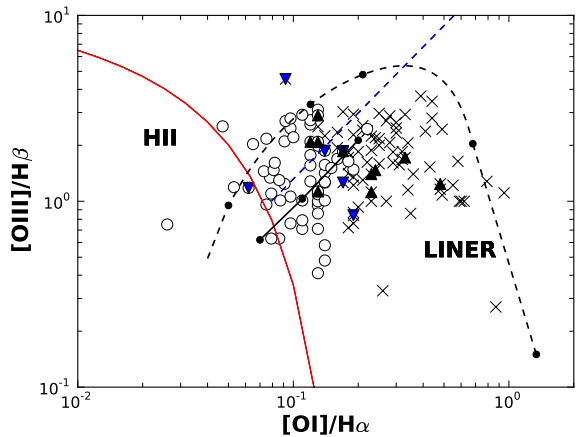}
\caption{Plot of [Oiii]/H$\beta$ vs [Oi]/H$\alpha$ for all L2(crosses)
 and T2 objects (open circles) in the sample of \citep{b16}. 
Highlighted are those objects (both L2 and T2) with 
(filled-in blue triangles pointing down) and without 
(filled-in black triangles) compact, flat-spectrum radio cores. Dashed line
denotes simulated optical line ratios from a 'low-hard' ULX continuum (from 
\citet{b12}), with the filled-in points on this curve corresponding to 
continuum starting energies of 1,1.5,2,5 and 10Ryd from left to right 
respectively (see text for details).The solid curve corresponds to a 
'soft-high' ULX continuum (from \citet{b12}), with the filled-in points 
corresponding to (from left to right) blackbody temperatures of 
$10,5,2\times 10^{6}$K respectively (see text for details).
\label{fig:o3o1}}
\end{figure}

In \citet{b98} we suggested that nuclear ULXs could power certain ('radio quiet') LINER 2s. 
In order to investigate the LINER 2 population, we extracted all the LINER 
2 (L2) and transition 2 (T2) objects from the Palomar sample of \citet{b16}. 
We included T2 objects since they occupy the same 'LINER' region of diagnostic 
diagrams. Figure~\ref{fig:o3o1} shows the optical diagnostic diagram of 
$[O_{III}]/\rm{H}_{\beta}$ vs $[O_{I}]/\rm{H}_{\alpha}$ for 
these objects, where L2s are denoted by crosses and T2s by open circles. Where radio data is available, we have
superimposed symbols on some of these objects. The 
blue, downward pointing triangles correspond to L2 and T2 objects 
\emph{without} flat-spectrum radio cores (or 'radio quiet'). Black, upward 
pointing triangles denote L2 and T2 objects \emph{with} flat-spectrum radio 
cores. The red curve corresponds to the starburst limit curve and the blue 
dashed line corresponds to the extreme mixing line (separating AGN and LINERs) 
\citep{b43,b44}. Corresponding diagrams for $[N_{II}]/\rm{H}_{\alpha}$ and 
$[S_{II}]/\rm{H}_{\alpha}$ look similar to Fig.~\ref{fig:o3o1} (see also 
\citet{b9}). Interestingly, the only distinction between the 'radio quiet' and 
'radio loud' populations of L2 and T2 objects seems to be that the radio quiet
 objects have $[O_{I}]/\rm{H}_{\alpha}<0.2$, implying FUV continuum 
penetrating relatively dense material, suggestive of star formation. 
\citet{b44} show that the optical line ratios for most LINERs could be 
produced by a 'mean' AGN SED with log $U\leq-3$ and slightly harder 
ionizing radiation fields. 

To investigate the possibility that ULXs or XRBs 
with log$U\sim -3.0$ could push line ratios out of the starburst region and 
into the transition/LINER region of the optical diagnostic diagrams, we ran 
simulations with CLOUDY v8.0, described 
by \citet{b77}. For the ULX continuum, we used X-ray band continuum values from 
the observational study of ULX populations in nearby galaxies in \citet{b12}. In this study,
spectra of all the brightest X-ray sources were extracted from a sample of 32 galaxies. ULX 
spectra were categorized based on their luminosity and spectral shape into 'low-hard' state 
sources, with an average power-law index of $\Gamma=2.1$, and 'soft-high' 
state sources, with a power-law index of $\Gamma=2.5$ plus a blackbody with 
temperatures in the range [0.1,1]keV. In AGN, the $[O_{III}]$ line strength 
requires a compact narrow line region, moderate density ($\sim 10^{3} 
\rm{cm}^{-3}$) and a moderate covering fraction (0.02-0.2) \citep{b63}, so we 
used these values as a guide. Therefore, our ULX continuum radiation was 
chosen to ionize a sphere of material with a filling factor of $10\%$ and a 
density of $\sim 10^{3}\rm{cm}^{-3}$ located at $10^{17}-10^{17.5}$cm from the
 continuum source. This corresponds to log$(N_{H})=20.3$ which is less than 
the fiducial log$(N_{H})=21$ for NLR clouds. The powerlaw continuum was 
assumed to span 1-1000Ryd (unless 
otherwise specified) and the powerlaw and blackbody in the 'soft-high' state 
were assumed to have identical ionization parameters such that total 
log $U=-3.0$. Note that a wide range of ionization parameter (log U) is likely for ULXs since 
this depends on a wide range of parameters (such as the density and location of ionized 
material, the nature of the black hole accretion and outflows). Since ULXs are comparably bright 
to low luminosity AGN in X-rays, we simply chose a value of log U appropriate to low lumnosity 
AGN. Future simulations will be required to understand the dependence of line ratios on a 
variable log U. In Fig.~\ref{fig:o3o1} the dashed (interpolated) curve 
corresponds to simulated line ratios produced by the 'low-hard' ULX spectrum, 
and filled-in points 
along the curve correspond to a power-law continuum spanning (from 
left to right) 1,1.5,2,5,10-1000Ryd. The shorter solid (interpolated) curve in 
Fig.~\ref{fig:o3o1} corresponds to simulated line ratios due to a 
'soft-high' ULX ionizing continuum, where filled-in points correspond to a 
blackbody at temperature (from left to right), $10,5,2\times 10^{6}$K. 

The first point to make from Fig.~\ref{fig:o3o1} is that simulated ULX 
continua seem perfectly 
capable of generating optical line ratios observed in transition-type objects 
and LINERs. The ULX ionizing continuum must include some FUV 
($\sim$1-10Ryd) in order to produce $[O_{I}]/H_{\alpha}$ ratios in the correct 
range, but apart from this the simulated line ratios do not depend very
strongly on the continuum shape. The second point from Fig.~\ref{fig:o3o1} is 
that ULX black hole mass can determine whether the optical line ratio is 
transition-like or LINER-like. For example, a blackbody temperature of 
$5\times 10^{6}$K, corresponds to emission from an accretion disk around 
a $\sim 10M_{\odot}$ black hole accreting near Eddington luminosity and 
generates a line ratio (middle point on solid curve) in the middle of the 
transition object region. By 
contrast, a blackbody temperature of $2\times 10^{6}$K corresponds to an 
accretion disk 
around a $\sim 10^{3}M_{\odot}$ black hole and generates a line ratio 
(rightmost point on solid curve) in the 
LINER region. From Fig.~\ref{fig:o3o1}, the 'radio quiet' L2 and T2s 
have $[O_{I}]/H_{\alpha}<0.2$, so if they are powered by 'soft-high' state
 ULXs as we suggested in \citet{b98}, the black hole masses must be 
$<10^{3}M_{\odot}$ and some FUV continuum is required (possibly from star 
formation). We will carry out more detailed simulations in the future to 
understand the limits on optical line ratios for different values of 
log U, the absorbing column and different ionizing continua, nevertheless 
for the purposes of this Letter, ULXs can in principle generate optical line
 ratios observed in LINER and transition-type nuclei.

\subsection{Observational Constraints}
\label{sec:obs}
Recently \citet{b22} claimed that nuclear X-ray emission in LINERs could not 
be due to high mass XRBs since populations of young stars in these nuclei are 
generally ruled out. Of course, as discussed above, it is not necessary to 
have populations of young stars to 
account for X-ray observations of low luminosity galactic nuclei. \citet{b17} 
find that X-ray/optical flux ratios for optical counterparts to 
ULXs are generally consistent with LMXB in \emph{old} clusters. Furthermore, 
in M31, the distribution of variable X-ray point sources in the innermost 
$\sim 450$pc may be consistent with an ageing population of low mass XRBs 
\citep{b36}. Therefore, integrating 
over the contributions from low mass XRBs is perfectly capable of powering 
nuclear X-ray emission for Gyrs and potentially generating a LINER-like or 
transition object-like appearance. 
Although low mass XRBs tend to be transient, they can actually dominate the 
XLF with reasonable choice of duty cycles \citep{b38}. For example, a choice 
of outburst rate (O.R.)$\sim 10\%$ during $\sim 75\%$ of the lifetime of 
XRBs is a reasonable estimate for the nucleus of Cen A \citep{b38}. 
\citet{b83} suggest O.R.$\sim 1\%$ for an estimated population of $\sim 
10-10^{3}$ binaries within $<1$pc of SgrA* could account for XRT observations. 
ULXs if unbeamed could have duty cycles as high as $\sim 10\%$ \citep{b85}.

We should expect (at 
least) several low mass XRBs in $\sim 0.5-1''$ X-ray observations of 
most galactic nuclei. Nuclear ULXs, like that in M82 should occur with 
moderate levels of star formation in the nucleus, although ULXs are observed 
in early-type galaxies at a rate of a few per galaxy \citep[e.g.][]{b41}. 
Indeed the mass of the ULX in M82 suggests that it is cannibalizing its host 
cluster or has captured companions. Another possible inconsistency between 
LINER X-ray emission and ULX or XRB emission is the XLF of LINER nuclei 
\citep{b22}. However, the sample size (82) of \citet{b22} is limited (see 
\citep{b23} for discussion of the dangers of this). Furthermore, their 
cumulative power-law indices ($\sim -0.2,-0.8$) 
before and after the power-law break are actually not that different from 
the \emph{cumulative} power-law indices of XRBs at low 
luminosities ($\sim -0.8$) in \citep{b23} or even from M31 \citep{b28}, 
although these steepen at higher luminosities. A much larger LINER sample is 
evidently required for a reliable understanding of the LINER XLF.

The UV band is important in our discussion, since emission from an accretion 
disk around supermassive black holes should peak in the UV band. \citet{b29} 
found evidence for UV variability in LINER 1s 
and LINER 2s. However of the five LINER 2s in \citet{b29} without a compact 
radio core, none varied at $>95\%$ confidence and three of the five 
radio-quiet LINER 2s 
(NGC 3486, NGC 4569 and NGC 5055 at 9,1 and 8Mpc distant respectively) were 
consistent with no UV variation 
whatsoever \citep{b29}. This suggests that hot stars rather than AGN 
are powering the UV emission in radio-quiet LINER 2s. X-ray imaging of 
LINERs reveals that extended emission or complex clumpy emission is common in 
LINER 2 nuclei. Extended emission can be explained in a nuclear ULX 
model by several ULXs/XRBs in a nucleus, in a region of hot massive stars. In 
a small sample of L2 and T2 nuclei, around half showed clear evidence for 
extended emission \citep{b27}. Extended 
emission may be present in the other L2 and T2 nuclei in the sample of 
\citet{b27} but these objects are faint. 

\subsection{Future Observational Constraints}
\label{sec:future}
Several observational approaches can be taken to isolate 
emission from minor black holes in L2 and T2 galactic nuclei. In the X-ray 
band, high angular resolution imaging and timing studies of nearby L2 and T2 nuclei 
will put limits on the number and accretion rate 
of candidate ULXs. In the optical band, variation in the optical 
line ratios of L2 and T2 nuclei will determine whether the source of the 
ionizing radiation is varying rapidly. Also in the optical band, high angular 
resolution studies of these nuclei will allow us to identify possible 
counterparts to sources of nuclear X-ray emission (for example, a nuclear 
globular cluster containing an accreting IMBH). In the IR band, high 
angular resolution images will identify star forming regions and their 
proximity to nuclear X-ray sources, as well as putting limits on 
reprocessed ionizing radiation that could originate from ULXs.

In this letter we have emphasized searches for accreting minor black holes in 
galactic nuclei without flat-spectrum radio cores. However, we note that 
stellar mass black 
holes also show evidence for radio jets, with flat or steep spectra depending 
on their X-ray state \citep[e.g.][]{b51}. Radio jets may also 
be associated with these ULXs \citep[e.g.][]{b50,b52}. So, some L2s and T2s 
with flat-spectrum radio jets may also be powered by ULXs, although this is 
beyond the scope of the present work. Finally, we briefly note that while 
X-ray bright accretion onto the central 
supermassive black hole will drown out emission from nearby minor black 
holes, signatures of minor black holes could still show up. Signatures might 
include a break in the nuclear X-ray luminosity function, a 
soft X-ray excess that varies on much faster timescales than 
the rest of the X-ray continuum and characteristic breaks in the 
power spectral density of the nucleus. 
\section{Conclusions}
\label{sec:conclusions}
We should expect large numbers of minor black holes in galactic nuclei as a 
consequence of stellar evolution, minor mergers and gravitational dynamical 
friction. If the central supermassive black hole is relatively quiescent, 
actively accreting minor black holes could dominate the nuclear X-ray emission.
This could result in the mis-classification of LINER and transition-type 
activity in galactic nuclei. Furthermore, inspiralling minor black holes could 
be important sources of gravitational radiation and if accreting, could be 
very useful in studying predictions of strong gravity. Here we show that 
nuclear ULXs are capable of producing the optical line ratios observed in 
LINER and transition-type objects. If nuclear ULXs are responsible for the 
activity in radio-quiet LINER 2 nuclei, as suggested in \citet{b98}, 
preliminary simulations indicate that the ionizing continuum is likely to be 
'soft-high' and originating from around black holes with mass 
$<10^{3}M_{\odot}$. We 
outline some future observational strategies to constrain emission from minor 
black holes in galactic nuclei.
\section*{Acknowledgements}
We acknowledge very useful discussions with M. Coleman Miller on dynamical 
friction and black hole mergers. BM \& KESF acknowledge support from CUNY 
grant CCRI-06-22.


\label{lastpage}

\end{document}